\documentclass[aps,twocolumn,showpacs,preprintnumbers,amsmath,amssymb,prl]{revtex4-1}
\usepackage{epsf,amsmath,amssymb,amsfonts,verbatim,color,multirow,pifont}
\usepackage{graphicx}

\begin{document}
\title{Threshold for everlasting initial memory in equilibration processes}
\author{J.S.~Lee$^{1}$}
\author{Chulan~Kwon$^{2}$} %\email{ckwon@mju.ac.kr}
\author{Hyunggyu~Park$^{1}$} %\email{hgpark@kias.re.kr}
\affiliation{{$^1$School of Physics, Korea Institute for
Advanced Study, Seoul 130-722, Korea}\\{$^2$Department of Physics, Myongji University, Yongin, Gyeonggi-Do 449-728, Korea}}

\date{\today}

\begin{abstract}
Conventional wisdom indicates that initial memory should decay away exponentially
in time for general (noncritial) equilibration processes. In particular,
time-integrated quantities such as heat are presumed to lose initial memory
in a sufficiently long-time limit. However,  we show that
the large deviation function of time-integrated quantities may exhibit
initial memory effect even in the infinite-time limit, if the system
is initially prepared sufficiently far away from equilibrium.
For a Brownian particle dynamics, as an example, we found a sharp finite threshold
rigorously, beyond which the corresponding large deviation function contains
everlasting initial memory. The physical origin for this phenomenon is
explored with an intuitive argument and also from a toy model analysis.

\end{abstract}


\maketitle
Hot coffee gets colder and iced coffee gets warmer at room temperatures.
These phenomena are the examples of equilibration processes and can be generalized
as the following situation; a system with initial temperature $T_s$ is in thermal contact
with a heat bath with temperature $T_b$. Then, the system gradually deviates from
its initial state and approaches to the final equilibrium (EQ) state which is determined
by the heat bath. Here, the initial distance from final
equilibrium is parameterized by the temperature ratio $\beta\equiv T_b/T_s$.
The relaxation process is usually exponentially fast, so
the memory of the initial temperature will be lost for average values of
most physical observables after a characteristic relaxation time.
However, the initial memory can often survive in the
tail part (rare-event region) of a probability distribution function (PDF)
even in the long-time limit.

What about {\em time-integrated} quantities such as heat, work, or
entropy production, which are the key quantities for nonequilibrium (NEQ)
fluctuation theorems~\cite{Evans,Gallavotti,Jarzynski,Kurchan,Lebowitz}?
These accumulated quantities are also affected
by a finite transient period, but their average values increase (or decrease)
linearly in time asymptotically in NEQ steady state.
Therefore, in a sufficiently long-time limit,
our conventional wisdom may lead us to expect that they will lose all initial memory (independent of $\beta$).
Nevertheless, in this Letter, we show rigorously that this is false wisdom
for time-integrated quantities
and, in particular, corresponding large deviation functions
depend strongly on the initial condition ($\beta$) even in the infinite-time limit.
More surprisingly, there exists a sharp threshold for $\beta^{-1}$
in general, only beyond which the initial memory lasts forever.

In literatures, there have been some reports that initial conditions
can affect the large deviation function in the long time
limit~\cite{Zon,Farago,Sabhapandit,Puglisi}.
For example, van Zon and Cohen~\cite{Zon} showed that heat transfer $Q$
in a driven harmonic oscillator in contact with a heat bath violates
the fluctuation theorem even in the long-time limit, starting initially from EQ. In contrast to work, heat is known to satisfy the fluctuation theorem, only starting with a uniform distribution
(infinite-temperature initial state)~\cite{hgpark1}.
Thus, their finding can be interpreted as an everlasting initial memory effect
in the large deviation function for heat.

In this Letter, we consider heat transfer
during the equilibration process of a simple Brownian particle and
investigate initial memory effects systematically in the long-time limit.
The Brownian particle dynamics is described by the Langevin equation
\begin{equation}
\dot{v} = -\gamma v + \xi, \label{Langevin}
\end{equation}
where $v$ is the velocity of the particle, $\gamma$ is the dissipative coefficient, and $\xi$ denotes a random white noise satisfying $\left< \xi(\tau) \xi(\tau^\prime) \right> = 2D\delta(\tau-\tau^\prime)$. Here, we set the particle mass $m=1$ for convenience and
the heat bath temperature $T_b = D/\gamma$. Initially, the system is prepared
in EQ state with the
Boltzmann distribution at temperature $T_s=T_b/\beta$. And then, the thermal contact is formed at time $\tau=0$ between the system and the heat bath, and maintained until final time
$\tau=t$.

Time-integrated heat flow between the system and the heat bath can be decomposed into the dissipated energy flow $Q_d$ from the system into the heat bath and the injected energy flow $Q_i$
in the other way around~\cite{Farago}:
\begin{equation}
Q_d \equiv \int_0^t d\tau~ \gamma v^2 ~~{\rm and}~~
Q_i \equiv \int_0^t d\tau~ \xi v.
\end{equation}
Even if the system reaches EQ in the long-time limit,
each of $\langle Q_d\rangle $ and $\langle Q_i\rangle$ increases linearly in time $t$ indefinitely with their difference
representing the system energy change
$\langle\Delta E\rangle=\frac{1}{2}[\langle v^2 (t)\rangle-\langle v^2(0)\rangle]$,
which is finite for nonzero $\beta$. As expected, there will be no net heat flow at EQ.

We first study the PDF of the (average) dissipated
power, $\varepsilon_d\equiv Q_d/t$, and later the injected power,
$\varepsilon_i\equiv Q_i/t$.
To calculate the PDF, $P(\varepsilon_d)$, it is convenient to consider its generating function
\begin{equation}
\pi_d (\lambda)=\left< e^{-\lambda t \varepsilon_d} \right>
= \int_{-\infty}^{\infty} d\varepsilon_d P(\varepsilon_d) e^{-\lambda t \varepsilon_d},
\label{generatingf}
\end{equation}
which is the Fourier transform of $P(\varepsilon_d)$.
The generating function can be calculated exactly by the standard path integral method~\cite{Farago,Wiegel}. With the initial Boltzmann distribution $P_{\rm init}(v(0))\sim \exp [-v^2(0)/(2 T_s)]$ at temperature
$T_s=D/(\gamma\beta)$, we find
\begin{equation}
\pi_d (\lambda) = e^{\gamma t/2} \left( \cosh \eta \gamma t + \frac{1+\widetilde{\lambda}/\beta}{\eta} \sinh \eta \gamma t \right)^{-1/2}, \label{pi_d}
\end{equation}
with dimensionless parameters $\widetilde{\lambda} = 2D \lambda/\gamma$ and $\eta = \sqrt{1+2\widetilde{\lambda}}$.

\begin{figure}
\centering
\includegraphics[width=1.00\linewidth]{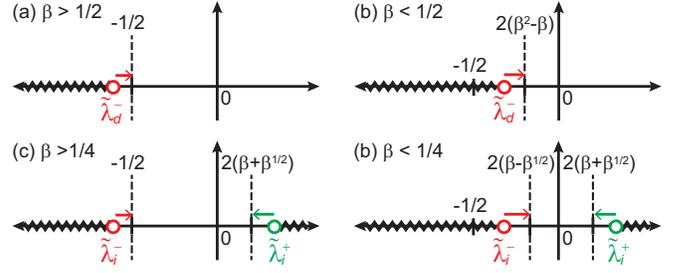}
\caption{(Color online) (a) and (b) show the branch-cut structure of $\pi_d (\lambda)$ on the complex $\widetilde{\lambda}$ plane for $\beta>1/2$ and $\beta<1/2$, respectively. (c) and (d) show the branch-cut structure of $\pi_i (\lambda)$ for $\beta>1/4$ and $\beta<1/4$, respectively. Wiggled lines denote branch cuts. A head of an each arrow locates at its asymptotic value of the respective branch point
as $t \rightarrow \infty$.}\label{branch}
\end{figure}

The inverse Fourier transform of Eq.~(\ref{pi_d}) yields the PDF
in terms of $\widetilde{\varepsilon}_d \equiv \varepsilon_d/D$ as
\begin{eqnarray}
P(\widetilde{\varepsilon}_d) &=& \frac{\gamma t}{4\pi i} \int_{-i\infty}^{i\infty}
d \widetilde{\lambda}~ \pi_d(\gamma\widetilde{\lambda}/2D) \exp\left[\frac{\gamma t \widetilde{\varepsilon}_d \widetilde{\lambda}}{2}\right] \nonumber \\
&=& \frac{\gamma t}{4\pi i} \int_{-i\infty}^{i\infty} d \widetilde{\lambda} \frac{ \exp\left[{\frac{\gamma t}{2}(\widetilde{\varepsilon}_d \widetilde{\lambda} + 1)}\right]}{ \sqrt{\cosh \eta \gamma t+ \frac{1+\widetilde{\lambda}/\beta}{\eta} \sinh \eta \gamma t} }. \label{P_d_integration}
\end{eqnarray}
For large $t$, the above integration can be carried out by the saddle point approximation. However, care should be taken due to the presence of the branch cut. Here, we take the
branch cut on the real-$\widetilde{\lambda}$ axis where
\begin{equation}
f \equiv \cosh \eta \gamma t + \frac{1+\widetilde{\lambda}/\beta}{\eta} \sinh \eta \gamma t \label{f}
\end{equation}
becomes negative, see Fig.~\ref{branch}.

We locate the branch points for large $t$, which depend on $\beta$.
Note that $\eta$ is real and positive for $\widetilde{\lambda}>-\frac{1}{2}$, while
$\eta$ becomes pure imaginary
for $\widetilde{\lambda}<-\frac{1}{2}$.
For $\beta>\frac{1}{2}$, $f>0$ and we have no branch points for $\widetilde{\lambda}>-\frac{1}{2}$.
Instead, we find them in the region of $\widetilde{\lambda}<-\frac{1}{2}$ and the largest
one is denoted by $\widetilde{\lambda}_{d}^-\simeq -\frac{1}{2}+ {\cal O}(t^{-2})$.
In the $t\to\infty$ limit, the branch point $\widetilde{\lambda}_{d}^-=-\frac{1}{2}$ has no
$\beta$ dependence. For $\beta<\frac{1}{2}$, in contrast, $f$ can become negative
for $\widetilde{\lambda}>-\frac{1}{2}$ and we find the branch point approaching
$\widetilde{\lambda}_{d}^-= -2\beta(1-\beta)$ as $t\to\infty$.
Locations of $\widetilde{\lambda}_{d}^-$'s and branch cuts are shown in
Figs.~\ref{branch}(a) and (b). It turns out that the branch-cut structure
plays a crucial role in determining the everlasting initial memory effect.

From Eq.~(\ref{P_d_integration}), one may easily expect
for large $t$
\begin{equation}
P(\widetilde{\varepsilon}_d)\simeq \exp\left[t h_t(\widetilde{\varepsilon}_d)\right]
\label{ldf}
\end{equation}
with the large deviation function (LDF)
$h(\widetilde{\varepsilon}_d)\equiv \lim_{t\to\infty}h_t(\widetilde{\varepsilon}_d)$.
We first calculate the LDF using the saddle point method in the presence of
the branch-cut structure found as above.
For large $t$, Eq.~(\ref{f}) becomes $f\simeq\frac{1}{2}e^{\eta \gamma t}\left( 1+\frac{1+\widetilde{\lambda}/\beta}{\eta}\right)$.
The saddle point $\widetilde{\lambda}_d^*$ is given by the solution of the following equation:
\begin{equation}
\frac{d}{d\widetilde{\lambda}}\left[\frac{\gamma t}{2}(\widetilde{\varepsilon}_d \widetilde{\lambda} +1-\eta)-\frac{1}{2}\ln \left( 1+\frac{1+\widetilde{\lambda}/\beta}{\eta}\right) \right] =0, \label{saddle_point_eq}
\end{equation}
where the logarithmic term is included because it may become very large
in the vicinity of $\widetilde{\lambda}=-2\beta(1-\beta)$ for $\beta<\frac{1}{2}$.

For $\beta>\frac{1}{2}$, we find a solution (saddle point) on the real-$\widetilde{\lambda}$
axis which is outside of the branch cut as
\begin{equation}
\widetilde{\lambda}_d^* = -\frac{1}{2}\left( 1-\frac{1}{\widetilde{\varepsilon}_d^2} \right),
\label{saddle_point_dissipated_1}
\end{equation}
as $t\to\infty$. Then, the LDF for $\beta>1/2$ becomes
\begin{equation}
h(\widetilde{\varepsilon}_d) =\frac{\gamma}{2}(\widetilde{\varepsilon}_d \widetilde{\lambda}_d^* +1 - \eta^*)= -\frac{\gamma }{4 \widetilde{\varepsilon}_d} (\widetilde{\varepsilon}_d-1)^2, \label{characteristic}
\end{equation}
where $\eta^* = \sqrt{1+2\widetilde{\lambda}_d^*}$ and
the logarithmic term is negligible. As $P(\widetilde{\varepsilon}_d)=0$ for
$\widetilde{\varepsilon}_d\le 0$, the LDF is defined only for $\widetilde{\varepsilon}_d> 0$.
This LDF has no $\beta$ dependence but is determined only by the heat bath properties ($\gamma$, $D$). Thus, we call Eq.~(\ref{characteristic}) the heat-bath characteristic curve (HBCC).

\begin{figure}
\centering
\includegraphics[width=1.00\linewidth]{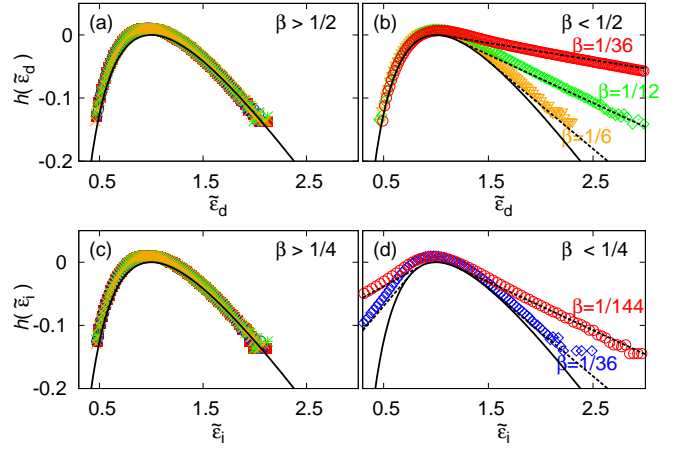}
\caption{(Color online) (a) and (b) are the LDF's of the dissipated power for $\beta>1/2$ and $\beta<1/2$, respectively. The solid line is the HBCC (see Eq.~(\ref{characteristic})). In (a) $\times$, $\ast$, $\circ$, and $\blacksquare$ are numerical data for $\beta=3/4$, $1$, $2$, and $4$, respectively. (c) and (d) are the LDF's of the injected power for $\beta>1/4$ and $\beta<1/4$, respectively. In (c) $\times$, $\ast$, $\circ$, and $\blacksquare$ are numerical data for $\beta=1/2$, $1$, $2$, and $4$, respectively. Each dashed line denotes the analytic line for each $\beta$. All numerical results are obtained at $t=100$.}\label{LDFs}
\end{figure}

For $\beta<\frac{1}{2}$,  the saddle point location exhibits a nonanalytic behavior as function of
$\widetilde{\varepsilon}_d$, due to the interplay of the saddle point and the branch point.
When $\widetilde{\varepsilon}_d < (1-2\beta)^{-1}$, the saddle point $\widetilde{\lambda}_d^*$
given by Eq.~(\ref{saddle_point_dissipated_1}) is located to the right of the branch point,
$\widetilde{\lambda}_d^*> \widetilde{\lambda}_d^-=-2\beta(1-\beta)$, in the $t\to\infty$ limit.
Thus, the LDF $h(\widetilde{\varepsilon}_d)$ is identical to the HBCC in
Eq.~(\ref{characteristic}). When $\widetilde{\varepsilon}_d > (1-2\beta)^{-1}$, the saddle point
approaches the branch point asymptotically from the right side, due to the divergence of the logarithmic term
in Eq.~(\ref{saddle_point_eq}) at the branch point. However, as this approach is not exponentially fast in time,
the dominant contribution to the LDF comes from the conventional first term in Eq.~(\ref{saddle_point_eq})
at the asymptotic saddle point $\widetilde{\lambda}_d^*= \widetilde{\lambda}_d^-=-2\beta(1-\beta)$.
Summarizing for $\beta<\frac{1}{2}$, the LDF is
\begin{eqnarray}
h(\widetilde{\varepsilon}_d) = \left\{
	\begin{array}{lc}
	-\frac{\gamma }{4 \widetilde{\varepsilon}_d} (\widetilde{\varepsilon}_d-1)^2, &  \widetilde{\varepsilon}_d < \frac{1}{1-2\beta}  \nonumber\\
	-\gamma \beta \left[ (1-\beta)\widetilde{\varepsilon}_d -1 \right], &  \widetilde{\varepsilon}_d > \frac{1}{1-2\beta}
	\end{array} \right. .\\\label{dissipated_LDF2}
\end{eqnarray}
Note that the LDF for large $\widetilde{\varepsilon}_d$ is deformed from the HBCC and has the initial condition ($\beta$) dependence, see Figs.~\ref{LDFs}(a) and (b).

Our results show that, for sufficiently high initial temperatures ($\beta<\frac{1}{2}$), the initial memory
survives forever in the large $\widetilde{\varepsilon}_d$ region of the LDF and
completely vanishes below the threshold of the initial temperature ($\beta>\frac{1}{2}$).
Large dissipated energy $Q_d$ is generated by the decay of highly energetic particles with energy $\sim Q_d$.
There are two distinct sources for highly energetic particles; $(a)$ heat bath and $(b)$ initial Boltzmann distribution,
which compete each other.
We estimate the probability $P_a$ and $P_b$
to find a particle to dissipate energy $Q_d$ from each source, respectively.
From the HBCC in  Eqs.~(\ref{ldf}) and (\ref{characteristic}), we find $P_a\sim \exp [-{Q_d}/(4 T_b)]$
for large $\widetilde{\varepsilon}_d$. On the other hand, we assume that
a particle with high initial energy decays by the deterministic dynamics 
of $\dot{v}=-\gamma v$. In this case, the dissipated energy is 
$Q_d=\frac{1}{2}v^2(0)$ in the long-time limit. 
Thus, we estimate $P_b\sim \exp [-{Q_d}/T_s]=\exp [-{\beta Q_d}/T_b]$ from the initial Boltzmann distribution at temperature $T_s$.
As a result, the HBCC $P_a$ dominates over $P_b$ for $\beta>\frac{1}{4}$ or 
the initial memory dominates, otherwise. 
As $Q_d$ is overestimated in the latter case,
the threshold value $\frac{1}{4}$ only sets its lower bound, which is
consistent with the correct value $\beta_d^c=\frac{1}{2}$.

We also calculate the leading finite-time correction of $h_t(\widetilde{\varepsilon}_d)$ in
Eq.~(\ref{ldf}). As the leading correction is ${\cal O}(\ln t /t)$, it yields a
power-law type prefactor to the exponential form of the PDF, $P(\widetilde{\varepsilon}_d)$. For $\beta<\frac{1}{2}$,
it is tricky to calculate this correction because the saddle point is very close
to the branch point. In fact, it cannot be obtained through a conventional Gaussian integral.
Here, we just report our result without presenting  details~\cite{Lee}
for $\beta<\frac{1}{2}$
\begin{equation}
P(\widetilde{\varepsilon}_d) =\left\{
\begin{array}{ll}
\frac{\sqrt{\gamma t}c_d(\beta)}{\widetilde\varepsilon_d\sqrt{(\widetilde\varepsilon_d+1)
((2\beta-1)\widetilde\varepsilon_d+1)}}
e^{-\frac{\gamma t }{4 \widetilde{\varepsilon}_d} (\widetilde{\varepsilon}_d-1)^2} & \textrm{(A)}\\
(\gamma t)^{3/4} r(\beta) e^{-\gamma\beta t \left[(1-\beta) \widetilde{\varepsilon}_d -1 \right]}
 & \textrm{(B)} \\
\frac{\sqrt{\gamma t}s(\beta)}{\sqrt{\widetilde{\varepsilon}_d - 1/(1-2\beta)} }
 e^{-\gamma\beta t \left[(1-\beta) \widetilde{\varepsilon}_d -1 \right]} & \textrm{(C)}
\end{array}
\right.
\label{pdf_2}
\end{equation}
where there are three regions: (A) $(1-2\beta)^{-1}-\widetilde{\varepsilon}_d\gg
(\gamma t)^{-1/2}$; (B) $|\widetilde{\varepsilon}_d-(1-2\beta)^{-1}|\ll(\gamma t)^{-1/2} $;
(C) $\widetilde{\varepsilon}_d- (1-2\beta)^{-1}\gg(\gamma t)^{-1/2}$. Three constants are given as $c_d(\beta)=\sqrt{{\beta}/{\pi}}$, $r(\beta)={\sqrt{2\beta}
(1-2\beta)^{7/4}\Gamma\left(\frac{1}{4}\right)}/({4\pi\sqrt{1-\beta}})$, and
$s(\beta)={\sqrt{\beta}(1-2\beta)}/{\sqrt{\pi(1-\beta)}}$.
For $\beta>\frac{1}{2}$, the PDF is given by the
same one in (A) of Eq.~(\ref{pdf_2}) for all $\widetilde{\varepsilon}_d>0$.
The prefactors depend on the initial condition ($\beta$) for all cases,
as expected, but their power-law exponent in terms of $\widetilde{\varepsilon}_d$
changes abruptly from $-2$ to $-\frac{1}{2}$ as
$\widetilde{\varepsilon}_d$ increases. It is interesting to note that
this exponent change is very similar to what was found
dynamically for the PDF of nonequilibrium work in simple linear diffusion systems~\cite{Chulan}.

Now, we turn to the injected power, ${\varepsilon}_i=Q_i/t$. The calculation method is
almost the same as before. We obtain the generating function of the injected power
as
\begin{eqnarray}
\pi_i (\lambda) = e^{\gamma t/2} \left( \cosh \eta \gamma t + \frac{1+\widetilde{\lambda}-\widetilde{\lambda}^2/2\beta}{\eta} \sinh \eta \gamma t \right)^{-1/2} .
\end{eqnarray}
Compared to Eq.~(\ref{pi_d}), there is only a parametric difference in the coefficient of
the hyperbolic sine term.
We can obtain the PDF of the dimensionless injected power, $P(\widetilde{\varepsilon}_i)$
with $\widetilde{\varepsilon}_i \equiv \varepsilon_i /D$, by performing the inverse
Fourier transform of $\pi_i(\lambda)$.

Similar to the case of the dissipated power, the branch points are determined
by the equation
\begin{equation}
0= \cosh \eta \gamma t + \frac{1+\widetilde{\lambda} -\widetilde{\lambda}^2/2\beta}{\eta} \sinh \eta \gamma t . \label{g}
\end{equation}
We find two relevant solutions of Eq.~(\ref{g}); one is on
the positive real axis, $\widetilde{\lambda}_{i}^+$, and the other is
on the negative real axis, $\widetilde{\lambda}_{i}^-$, as shown in
Figs.~\ref{branch}(c) and (d), respectively.
In the $t\to\infty$ limit,
one can show that $\widetilde{\lambda}_{i}^+ = 2(\beta+\sqrt{\beta})$
for all $\beta$, while
$\widetilde{\lambda}_{i}^-=-\frac{1}{2}$
for $\beta>\frac{1}{4}$ and $\widetilde{\lambda}_{i}^-= -2(\sqrt{\beta}-\beta)$
for $\beta<\frac{1}{4}$.

By defining the LDF, $h(\widetilde{\varepsilon}_i)$, for  $\widetilde{\varepsilon}_i$
in the $t\to\infty$ limit and through a similar algebra, we find for $\beta>\frac{1}{4}$
\begin{eqnarray}
h(\widetilde{\varepsilon}_i) = \left\{
	\begin{array}{lc}
	-\gamma \sqrt{\beta} \left[
1-(1+\sqrt{\beta})\widetilde{\varepsilon}_i \right], &  \widetilde{\varepsilon}_i < \frac{1}{1+2\sqrt{\beta}}  \nonumber\\
	-\frac{\gamma }{4 \widetilde{\varepsilon}_i} (\widetilde{\varepsilon}_i-1)^2, &  \widetilde{\varepsilon}_i > \frac{1}{1+2\sqrt{\beta}}
	\end{array} \right. .\\\label{injected_LDF1}
\end{eqnarray}
Note that  $h(\widetilde{\varepsilon}_i)$ is defined for all
$\widetilde{\varepsilon}_i$. The negative tail of $P(\widetilde{\varepsilon}_i)$
is affected by the initial condition and
the non-analyticity of   $h(\widetilde{\varepsilon}_i)$
is present even in the EQ process at $\beta=1$.
For $\beta<\frac{1}{4}$, the LDF becomes
\begin{eqnarray}
h(\widetilde{\varepsilon}_i) = \left\{
	\begin{array}{ll}
	-\gamma \sqrt{\beta} \left[
1-(1+\sqrt{\beta})\widetilde{\varepsilon}_i \right], &\widetilde{\varepsilon}_i < \frac{1}{1+2\sqrt{\beta}}\\
	-\frac{\gamma }{4 \widetilde{\varepsilon}_i}(\widetilde{\varepsilon}_i-1)^2, \quad\frac{1}{1+2\sqrt{\beta}}< &\widetilde{\varepsilon}_i <\frac{1}{1-2\sqrt{\beta}}  \\
	 -\gamma \sqrt{\beta} \left[
(1-\sqrt{\beta})\widetilde{\varepsilon}_i -1\right],  &\widetilde{\varepsilon}_i > \frac{1}{1-2\sqrt{\beta}}
	\end{array} \right. \label{injected_LDF2}
\end{eqnarray}
Our results read that the negative tail always depends on the initial condition, but
the positive tail shows the threshold at $\beta_i^c=\frac{1}{4}$ where the initial
condition dependence starts to appear. Note that the threshold value
varies with the quantity interested. The leading finite-time correction is rather
complicated, which will appear elsewhere~\cite{Lee}.

To confirm our analytic calculations in Eqs.~(\ref{characteristic}),
(\ref{dissipated_LDF2}), (\ref{injected_LDF1}), and (\ref{injected_LDF2}),
we performed numerical integrations
of the Langevin equation, Eq.~(\ref{Langevin}). Here, we set $\gamma= D=1$,
and integration time interval $\Delta t = 10^{-3}$.
Figure~\ref{LDFs}(a) displays numerical data for $h(\widetilde{\varepsilon}_d)$ at $t=100$
for various values of $\beta>1/2$. Regardless of $\beta$, all numerical results
collapse well onto the HBCC as expected from Eq.~(\ref{characteristic}). Slight deviation from the analytic HBCC comes from the finite-time effect. We confirmed that the
LDF with leading finite-time correction (see Eq.~(\ref{pdf_2})) perfectly agrees with the numerical data at $t=100$ (not shown here). Figure~\ref{LDFs}(b) shows $h(\widetilde{\varepsilon}_d)$ for $\beta<1/2$. Numerical results also agree well with our analytic results in Eq.~(\ref{dissipated_LDF2}).
Figure~\ref{LDFs}(c) and (d) show the LDF of the injected power for $\beta>1/4$ and $\beta<1/4$, respectively. In Fig.~\ref{LDFs}(c) the LDF for $\widetilde{\varepsilon}_i<(1+2\sqrt{\beta})^{-1}$  does not appear simply
because the region is outside of the plot range. Meanwhile, the three regions are clearly seen in Fig.~\ref{LDFs}(d), as expected from Eq.~(\ref{injected_LDF2}).

\begin{figure}
\centering
\includegraphics[width=1.00\linewidth]{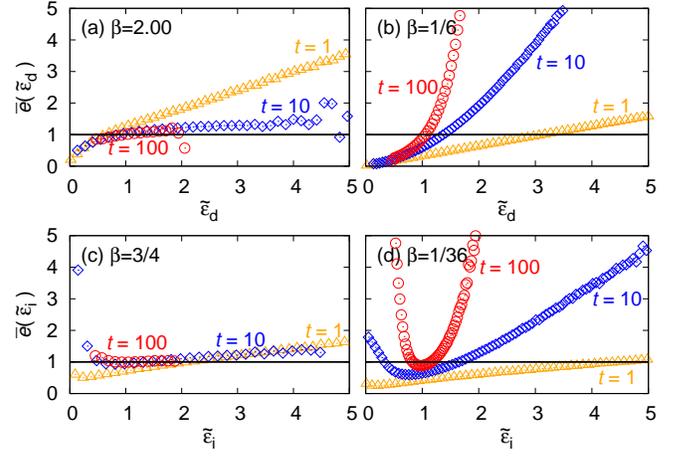}
\caption{(Color online) (a) and (b) show the correlation in time between the initial energy and the dissipated power for $\beta>1/2$ and $\beta<1/2$, respectively. (c) and (d) show the correlation between the initial energy and the injected power at time $\tau$ for $\beta>1/4$ and $\beta<1/4$, respectively. $\vartriangle$, $\lozenge$, and $\circ$ are data for $\tau=1$, $10$, and $100$, respectively.}\label{IEs}
\end{figure}

To understand better the origin of the threshold $\beta^c$ dividing
different phases, we introduce a simple toy model.
In order to examine the correlation between the initial energy
and the average power, we define a function
$\overline{E}_t (\widetilde{\varepsilon})$ which is the average initial energy of a particle whose average (dissipated or injected) power until time $t$ is
given by $\widetilde{\varepsilon}$. It is convenient to use the normalized function
$\overline{e}(\widetilde{\varepsilon})\equiv \overline{E}(\widetilde{\varepsilon})/\langle E \rangle_0$, with $\langle E\rangle_0$ the average initial energy without
any constraint on its power. In Fig.~\ref{IEs}(a), $\overline{e}(\widetilde{\varepsilon}_d)$
approaches the constant $1$ for large $t$, which implies no correlation between the initial
energy and the corresponding dissipation power. Thus, there will be no initial condition
dependence on the PDF for large $\widetilde{\varepsilon}_d$. In contrast,
 Fig.~\ref{IEs}(b) shows divergence of $\overline{e}(\widetilde{\varepsilon}_d)$ in time,
 which indicates that high initial energy is responsible for large $\widetilde{\varepsilon}_d$. Thus, the tail of the PDF, $P(\widetilde{\varepsilon}_d)$,
 should be dominated by a particle with high initial energy which is generated by
 the initial high-temperature distribution.
 This causes the deviation of $h(\widetilde{\varepsilon}_d)$
 from the HBCC for $\widetilde{\varepsilon}_d>(1-2\beta)^{-1}$ when $\beta<\frac{1}{2}$
 (higher initial temperatures). Figure~\ref{IEs}(c) shows $\overline{e}(\widetilde{\varepsilon}_i)$ approaches $1$ for $\beta>1/4$, while
 Figure~\ref{IEs}(d) shows its divergence for $\beta<1/4$.

In summary, we consider the equilibration process of a Brownian particle system,
staring from various initial temperatures different from the heat bath temperature.
We calculate the LDF of time-integrated quantities like the dissipated energy and
the injected energy due to the heat bath. Remarkably, we find a finite threshold
for the initial temperature, only beyond which the LDF contains everlasting initial
memory. We argue that this is due to the competition of
highly energetic particles originated from the heat bath and
from the initial distribution.
Our simple toy model analysis supports this argument by
showing that large dissipated energy is generated dominantly by
particles with high initial energy, rather than by highly energetic particles
randomly generated by the heat bath, when the initial temperature is sufficiently
high enough with respect to the heat bath temperature.
We expect that our results are applicable to general equilibration or nonequilibrium
processes, which implies that the rare-event measurements for time-integrated
quantities should be carefully carried out because the initial memory may
survive forever.

This research was supported by the NRF grant No.~2011-35B-C00014 (JSL)
and by Mid-career Researcher Program through NRF grant No.~2010-0026627
(CK,HP) funded by the MEST .

\vfil\eject
\end{document}